\documentclass[reprint, secnumarabic, amssymb, amsmath, superscriptaddress, aps, prl]{revtex4-2}
\usepackage{graphicx}
\usepackage[colorlinks=true,urlcolor=blue]{hyperref}
\usepackage{hyperref}
\usepackage{xcolor}
\hypersetup{
colorlinks,
linkcolor={red!50!black},
citecolor={blue!50!black},
urlcolor={blue!80!black}
}
\usepackage[normalem]{ulem}
\usepackage{units}
\begin{document}

\title{Anharmonic multiphonon origin of the valence plasmon in SrTi$_{1-x}$Nb$_x$O$_3$}

\author{Caitlin S. Kengle}
\author{Samantha I. Rubeck}
\author{Melinda Rak}
\author{Jin Chen}
\author{Faren Hoveyda}
\author{Simon Bettler}
\author{Ali Husain}
\thanks{Present address: Department of Physics and Astronomy and Quantum Matter Institute, University of British Columbia, Vancouver, BC V6T 1Z4, Canada}
\affiliation{Department of Physics, University of Illinois, Urbana, Illinois 61801}
\affiliation{Materials Research Laboratory, University of Illinois, Urbana, Illinois 61801}

\author{Matteo Mitrano}
\thanks{Present address: Department of Physics, Harvard University, Cambridge, MA 02138}
\affiliation{Department of Physics, University of Illinois, Urbana, Illinois 61801}
\affiliation{Materials Research Laboratory, University of Illinois, Urbana, Illinois 61801}

\author{Alexander Edelman}
\author{Peter Littlewood}
\affiliation{James Franck Institute, 929 East 57th Street, Chicago, Illinois 60637}
\affiliation{Department of Physics, The University of Chicago,5720 South Ellis Avenue, Chicago, Illinois 60637}
\author{Tai-Chang Chiang}
\author{Fahad Mahmood}

\author{Peter Abbamonte}
\affiliation{Department of Physics, University of Illinois, Urbana, Illinois 61801}
\affiliation{Materials Research Laboratory, University of Illinois, Urbana, Illinois 61801}
\email{abbamonte@mrl.illinois.edu}

\begin{abstract}

Doped SrTi$_{1-x}$Nb$_x$O$_3$ exhibits superconductivity and a mid-infrared optical response reminiscent of copper-oxide superconductors. Strangely, its plasma frequency, $\omega_p$, increases by a factor of $\sim$3 when cooling from 300K to 20K, without any accepted explanation. Here, we present momentum-resolved electron energy loss spectroscopy (M-EELS) measurements of SrTi$_{1-x}$Nb$_x$O$_3$ at nonzero momentum, $q$. We find that the 
IR feature previously identified as a plasmon is present at large $q$ in insulating SrTiO$_3$, where it exhibits the same temperature dependence and may be identified as an anharmonic, multiphonon background. Doping with Nb increases its peak energy and total spectral weight, drawing this background to lower $q$ where it becomes visible in IR optics experiments. We conclude that the ``plasmon" in doped SrTi$_{1-x}$Nb$_x$O$_3$ is not a free-carrier mode, but a composite excitation that inherits its properties from the lattice anharmonicity of the insulator.

\end{abstract}

\maketitle


The cubic perovskite SrTiO$_3$ is a quantum paraelectric that, when doped with niobium or oxygen, becomes a polaronic metal with dilute superconductivity \cite{ Muller1979, Collignon2019a, Takada1980, Ruhman2016, Gorkov2017}. Doped SrTiO$_3$ exhibits evidence for quantum criticality, a peculiar mid-infrared (IR) optical response, and violates the Mott-Ioffe-Regel limit (i.e., exhibiting so-called “bad metal” behavior), suggesting close parallels with the copper-oxide superconductors \cite{Edge2015, BEDNORZ1988, Collignon2019a, Lin2017}. SrTiO$_3$ and its doped variants therefore remain of perennial importance. 

A perplexing property of doped SrTiO$_3$ is that its plasma frequency, $\omega_p$ is temperature dependent. For example, in SrTi$_{1-x}$Nb$_x$O$_3$, $\omega_p$ increases by a factor of $\sim$3 as the material is cooled from 300 K to 100 K \cite{Gervais1993, Bi2006}. This change is normally interpreted as a changing effective mass of the conduction electrons, $m^*$, which appears to be supported by some transport experiments \cite{VanDerMarel2011, Collignon2020}. However, the reason why $m^*$ would change so much has been unclear. The leading explanation was the mixed polaron theory proposed by Eagles \cite{Eagles1996}, which postulates that the material contains both large and small polarons whose relative population changes with temperature, resulting in an apparent change in $m^*$. However, subsequent optics studies found that this picture is inconsistent with the total spectral weight in the Drude response as well as the size of the coupling constant, $\alpha \sim 2$, which do not allow for the presence of small polarons \cite{VanMechelen2008,Devreese2010}. The origin of the temperature-dependent plasma frequency in doped SrTiO$_3$ therefore remains unexplained. 

Here we report an energy- and momentum-resolved study of the dynamic charge response of the normal state of niobium-doped SrTiO$_3$ using momentum-resolved electron energy-loss spectroscopy (M-EELS) \cite{Vig2017}, with a focus on the momentum dependence of the IR plasmon.

\begin{figure}[b]
	\centering
	\includegraphics[width=1.0\linewidth]{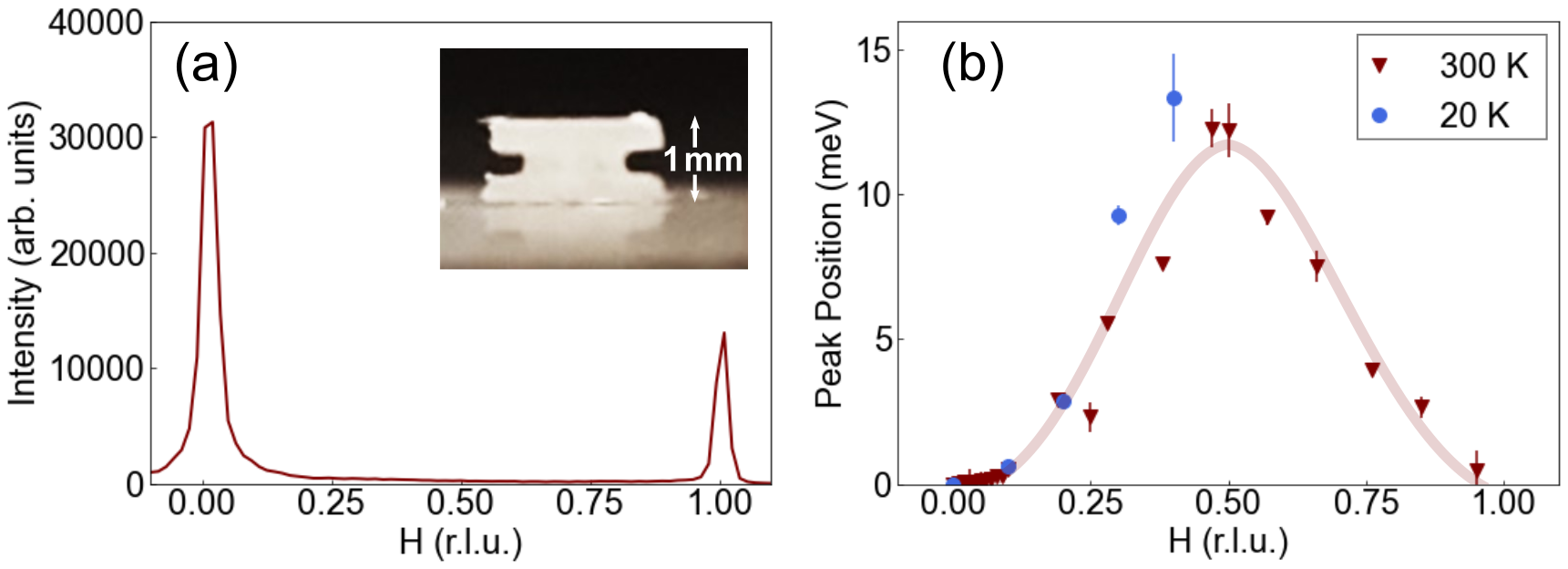}
	\caption{Room temperature characterization of sample surfaces (a) LEED scans along (H,0) showing the crystallinity of samples studied. Inset: notched undoped SrTiO$_3$ sample before fracturing. (b) Dispersion of peak of the TA phonon at room and base temperature. The line is a guide to the eye.}
	\label{fig:fig1}
\end{figure}

M-EELS measures the surface density-density correlation function, $S(q,\omega)$, its cross-section being given by

\begin{equation}\label{eq:1}
	\frac{\partial^{2}\sigma}{\partial\Omega \partial E} = \sigma_{0} M^2(q) \cdot S\left(q,\omega \right),
\end{equation}

\noindent where $\mathrm{\sigma_{0}}$ depends on the beam energy and surface reflectivity and $M(q) = 4\pi e^2/[q^2 + (k_i^z + k_s^z)^2]$ is the Coulomb matrix element \cite{Kogar2014, Vig2017}. The surface density response, $\chi^{\prime\prime}(q,\omega)$, is related to $S(q,\omega)$ by the fluctuation-dissipation theorem \cite{Kogar2014,Vig2017} and provides information about the collective modes of a material.


The correlation function at finite momentum, $S(q,\omega)$, was obtained by dividing $\mathrm{\sigma_{0}}$ and $M^2(q)$ from the raw data, which results in a modest correction to the shape of the spectra. For $q = 0$, $M^2(q)$ is a very rapidly varying function of $\omega$, which departs from the analytic form because of resolution effects. For this reason, at $q=0$ we just show the raw data. All spectra were normalized to the total spectral weight to place them on a similar scale for plotting purposes. 

M-EELS measurements were performed on samples of $\mathrm{SrTi_{1-x}Nb_{x}O_{3}}$ with $x$ = 0, 0.002, 0.01, and 0.014. Hall measurements of the $x$ = 0.002, 0.01, 0.014 samples showed them to be electron doped with carrier densities $(2.89\pm0.26)\times10^{19}$, $(1.49\pm0.14)\times10^{20}$, and $(2.13\pm0.07)\times10^{20}$ $\mathrm{cm^{-3}}$, respectively. Undoped samples with $x=0$ were too insulating for Hall measurements. So we take them to have zero carrier density. 

Surfaces were prepared by notching the sides of the samples (Fig.~\ref{fig:fig1}(a), inset) and fracturing in ultrahigh vacuum. See \cite{supp} for more information. These surfaces are of sufficient quality to achieve momentum conservation in M-EELS, demonstrated by resolution-limited (0,0) and (1,0) Bragg peaks (Fig.~\ref{fig:fig1}(a)) and a dispersing transverse acoustic (TA) phonon (Fig.~\ref{fig:fig1}(b)). This demonstrates that momentum is conserved in the current measurements and that M-EELS should reveal the dispersion of other collective modes. 

\begin{figure}
	\centering
	\includegraphics[width=1.0\linewidth]{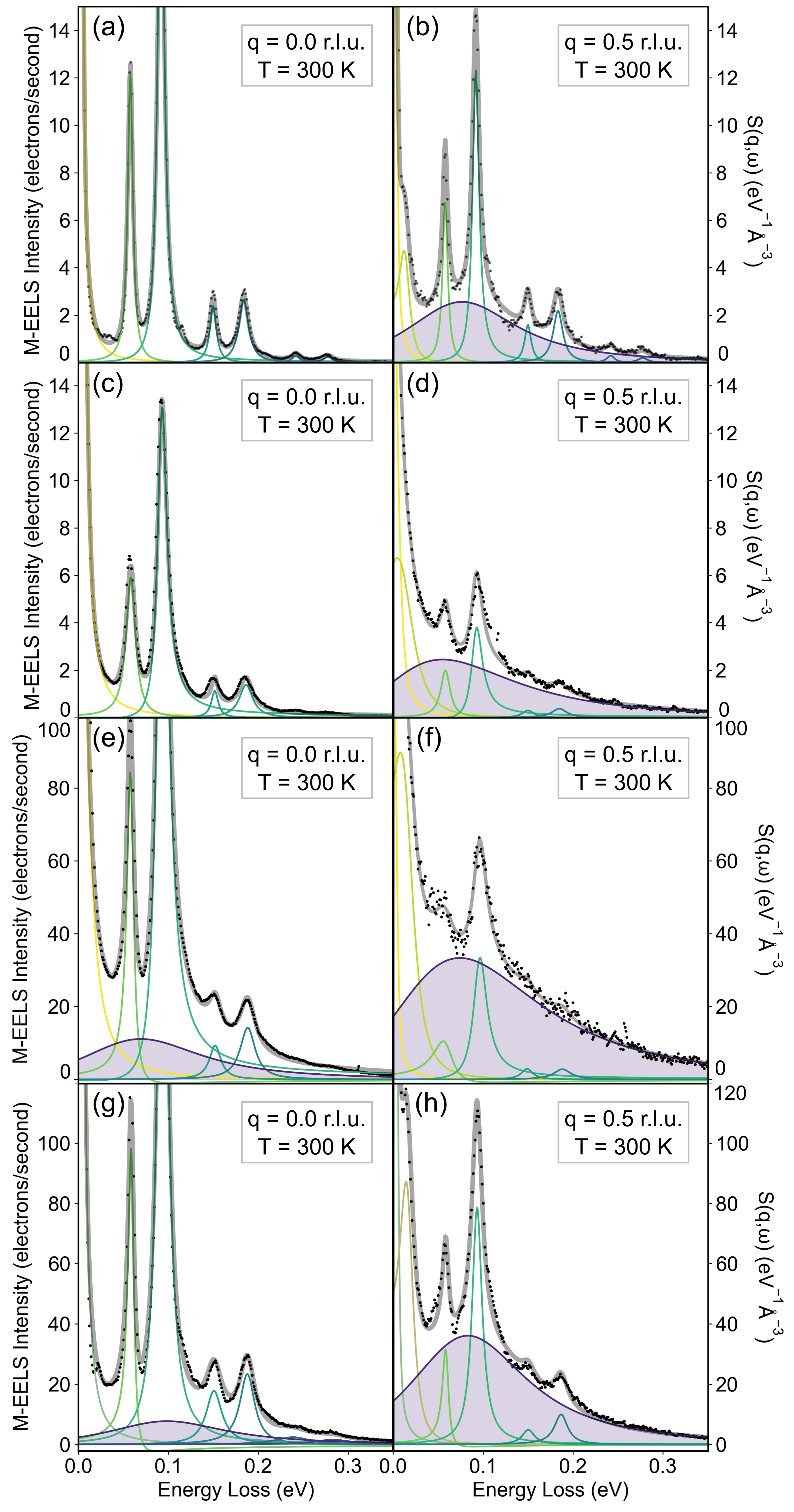}
	\caption{M-EELS Spectra at 300 K for $x = 0$ at (a) $q = 0$ and (b) $q = 0.50$ r.l.u., $x = 0.002$ at (c) $q = 0$ and (d) $q = 0.50$ r.l.u. $x = 0.01$ at (e) $q = 0$ and (f) $q = 0.50$ r.l.u., and $x = 0.014$ at (g) $q = 0$ and (h) $q= 0.50$ r.l.u. Solid lines represent fits to the phonons and the elastic line, while the shaded areas represent fits to the plasmon/multiphonon background.}
	\label{fig:fig2}
\end{figure}

We begin by discussing the insulator ($x=0$). At 300 K, the spectrum at the Brillouin zone center ($q = 0$ r.l.u.; Fig.~\ref{fig:fig2}(a)), shows a series of optical phonons and overtones, consistent with previous surface EELS studies \cite{Li2018, Conard1993}. With increasing momentum transfer, $q$, a TA phonon disperses out from the elastic line to a maximum frequency of $\sim$14 meV, consistent with previous neutron scattering studies \cite{Shirane1969, Choudhury2008, He2020}. 

At the Brillouin zone boundary, $q = 0.5$ r.l.u., a multiphonon background is visible that is absent at $q=0$ (Fig.~\ref{fig:fig2}(b)). 
This feature, which is common in anharmonic crystals \cite{Ashcroft76}, arises from scattering events involving two or more phonons in the final state. Because only the total momentum is conserved in scattering, processes of this sort do not constrain the momenta of the individual phonons, resulting in a lineshape that resembles the momentum-integrated phonon density of states . 

To quantify the results, we fit our data using a model comprising a pseudo-Voigt function for the elastic line, Fano functions for the optical phonons, a damped oscillator for the TA phonon, and a Drude function for the multiphonon background \cite{supp}. There is no physical reason, at this stage, for choosing a Drude form for the background, which is unrelated to metallic electrons. But this choice will become more meaningful when we analyze doped materials. The fit function was multiplied by a Bose factor before fitting, so the model can be considered to represent the susceptibility, $\chi^{\prime\prime}(q,\omega)$, rather than the correlation function, $S(q, \omega)$. This removes the effects of temperature on the population of excitations, enabling more meaningful comparison between spectra at different temperatures. The fits for the $x=0$ sample at $q = 0$ and 0.50 r.l.u. are shown in Fig. \ref{fig:fig2}(a),(b) and yield the TA phonon dispersion in Fig. \ref{fig:fig1}(b) (Data at all $q$ values may be found in the Supplement \cite{supp}).

\begin{figure*}
	\centering
	\includegraphics[width=1.0\linewidth]{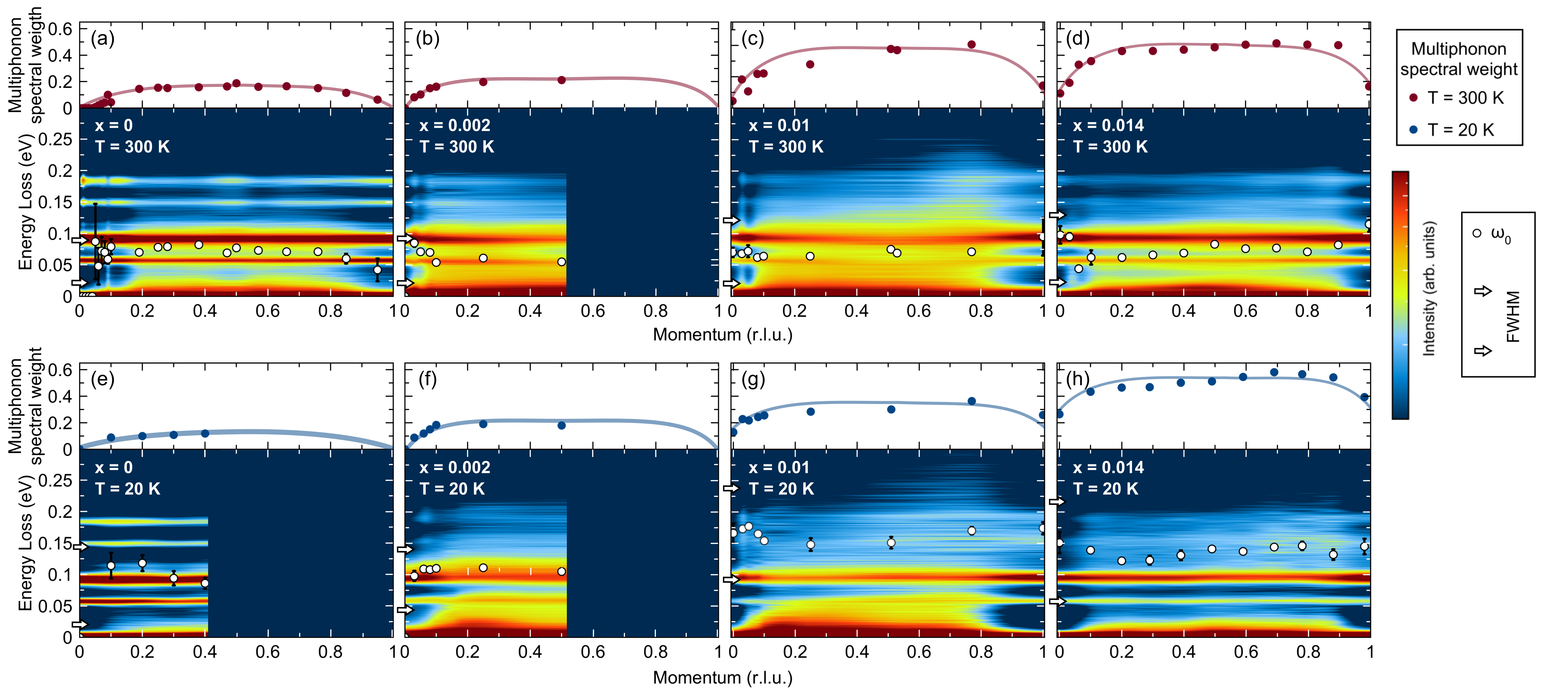}
	\caption{Full data sets taken at 300 K for (a) x = 0, (b) x = 0.002, (c) x = 0.01, and (d) x = 0.014 samples and at 20 K for (e) x = 0, (f) x = 0.002, (g) x = 0.01, and (h) x = 0.01. The overlaid points show the peak position of the continuum obtained from fitting. Arrows on the left of each panel indicate the full width half maximum of the continuum obtained from fitting. The plots above each panel show the integrated intensity of the continuum normalized by the total spectral weight. The solid red/blue curves are a guide to the eye. The peak position shifts up with increased doping and with decreased temperature. The integrated intensity of the continuum, normalized to the total spectral weight, increases with doping and decreases with temperature. 
	} 
	\label{fig:fig3}
\end{figure*}

We focus on the behavior of the multiphonon background, summarized in Fig.~\ref{fig:fig3}(a),(e). Our fits reveal that the background is absent at $q=0$ r.l.u., but becomes visible for $q \geq 0.08$ r.l.u. and grows with increasing $q$, reaching its maximum intensity at the Brillouin zone boundary. The lineshape of the background, characterized by its center frequency, $\omega_0$, and its full-width half-maximum, is $q$-independent. This is consistent with the notion that a multiphonon background is determined by the $q$-integrated density of states \cite{Ashcroft76}, though its coupling to the probe electron depends on the value of $q$. 

\begin{figure}
	\centering
	\includegraphics[width=1.0\linewidth]{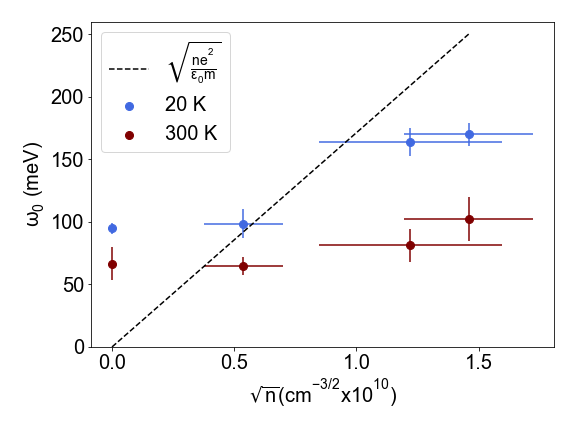}
	\caption{$\omega_0$ as a function of carrier concentration at room temperature and base temperature. Horizontal error bars represent errors in Hall conductivity measurements. Vertical error bars represent the first standard deviation in peak position between momenta. The dashed black line represents the plasma frequency calculated using a standard Drude form.}
	\label{fig:fig4}
\end{figure}

When cooled from $T=300$ K to $T=20$ K, the integrated intensity of the multiphonon background decreases (Fig.~\ref{fig:fig3}(a) versus (e), inset)
and its center frequency, $\omega_0$, increases from 66 meV to 95 meV (Fig.~\ref{fig:fig3}(a) versus (e), Fig.~\ref{fig:fig4}). This effect may be understood as a consequence of anharmonicity: in the presence of phonon-phonon interactions, the phonon self-energy depends on how many other phonons are present in the system. Strikingly, temperature dependence and lineshape of this background is similar to that of the IR plasmon in doped SrTi$_{1-x}$Nb$_x$O$_{3}$ \cite{Bi2006,Gervais1993,VanMechelen2008}, though our data were taken from an undoped, insulating sample in which no carriers are present. 

We now examine how the loss spectra change as the system is electron-doped. We consider first the most lightly doped SrTi$_{1-x}$Nb$_{x}$O$_3$ sample with $x = 0.002$, corresponding to one added charge carrier per $\sim 500$ unit cells. This sample is conducting, so one may expect a plasmon to be present, though polaronic effects may be strong. 

At 300 K and $q = 0$ r.l.u.~(Fig.~\ref{fig:fig2}(c)), the spectra look similar to that of the insulator, but with slightly higher background between the optical phonon peaks. At this doping, the phonons are significantly broader. In particular, the width of the 93 meV Fuchs-Kliewer optical phonon is a factor of 2.12 larger than at $x = 0$ (see \cite{supp}). This effect may be due to decay of the phonon into the elevated background at this composition. 

The intensity of the background increases with increasing $q$, while its lineshape remains unchanged, similar its behavior at $x=0$. By $q = 0.50$ r.l.u. the background is very pronounced and resembles the multiphonon continuum (Fig.~\ref{fig:fig2}(d)) in the insulator at the same $q$, though its intensity is higher than at $x=0$ (Fig.~\ref{fig:fig3}(a) versus (b)).

The background again fits well to a Drude form (Fig.~\ref{fig:fig2}(d)), but now contains significant weight for $q \geq$ 0.03 r.l.u., as compared to $q \geq$ 0.08 r.l.u. in the insulator. At this very light doping, the effect of the additional electrons is not to create a distinct Drude plasmon, but to increase the total spectral weight in the multiphonon background, and draw this weight to lower values of $q$. 

Changing the temperature has similar effects to the $x=0$ case. When cooled from $T=300$ K to $T=20$ K, 
the intensity of the multiphonon background decreases (Fig.~\ref{fig:fig3}(b) versus (f)), and $\omega_0$ increases from 64 meV to 98 meV, shown also in Fig.~\ref{fig:fig4}. This behavior is similar to the frequency shift of the plasmon seen in IR reflectivity experiments on materials with much higher niobium content \cite{Gervais1993,Bi2006,VanMechelen2008}.  

Further increasing the doping, to $x=0.01$, we find that the background is now clearly visible at $q = 0$ r.l.u. (Fig.~\ref{fig:fig2}(e)). 
The background has a similar energy and width as the feature identified as the plasmon in IR reflectivity measurements at this composition \cite{Gervais1993,Bi2006}, though we find that it evolves from the multiphonon background in the insulator and not purely from the addition of charge carriers. 

At this composition, the linewidths of the optical phonons are sharper than at $x = 0.002$ (though still broader than at $x=0$ \cite{supp}). We attribute this to a weakening of the electron-phonon interaction due to increased screening from additional charge carriers. Though the multiphonon background is stronger at this doping than at $x=0.002$, its coupling to the phonons is reduced, resulting in sharper phonon linewidths. 

As in the $x=0$ and $x=0.002$ samples, the intensity of the multiphonon background at $x=0.01$ increases with increasing $q$, while its lineshape remains unchanged (see Fig.~\ref{fig:fig3}(c)). The intensity of the background is higher than in either the $x=0$ or $x = 0.002$ samples (Fig.~\ref{fig:fig3}(c,d)). The center frequency, $\omega_0$, is also higher than in the other compositions, suggesting the multiphonon continuum is now acquiring some of the properties of a plasmon. 

The temperature-dependence of the multiphonon background at $x = 0.01$ exhibits the same behavior as in the other compositions. The overall intensity decreases, and $\omega_0$ increases from 81 meV to 164 meV as temperature is cooled from $T=300$ K to $T=20$ K (Fig.~\ref{fig:fig3}(c,g), Fig.~\ref{fig:fig4}). This shift is consistent with the plasmon reported in previous IR experiments at this composition \cite{Gervais1993,Bi2006,VanMechelen2008}. 

Turning now to the highest doping, $x=0.014$, the multiphonon background is clearly visible at $q=0$ r.l.u. (Fig.~\ref{fig:fig2}(g)) and has significant plasmon character. 
At $T=300$ K, both the intensity and $\omega_0$ (Fig.~\ref{fig:fig2}(g),(h)) are the highest of any of the dopings measured. 
Like in the other samples, the intensity of this excitation grows with increasing $q$ while its lineshape remains momentum-independent (Fig. \ref{fig:fig3}(d)). The optical phonons further sharpen (Fig. \ref{fig:fig2}(g),(h)), adding support to the notion that additional charge carriers lead to improved screening, further weakening the electron-phonon interaction.

Looking at the temperature dependence, we see a departure from previous trends. Cooling from $T=300$ K to $T=20$ K, the intensity of the multiphonon-plasmon now slightly increases, suggesting an enhancement of the susceptibility, $\chi''(q,\omega)$, at low temperature at this composition.
Nevertheless, $\omega_0$ exhibits the same behavior as the other dopings, increasing from 100 meV to 171 meV as the system is cooled from $T=300$ K to $T=20$ K, shown in Fig.~\ref{fig:fig4}. 
 
The overall trend suggests that the excitation identified as a Drude plasmon in IR optics \cite{Gervais1993,Bi2006,VanMechelen2008,Devreese2010} is not a simple, free carrier mode. Rather, it originates from multiphonon effects in the insulator that take on some of the characteristics of a plasmon as carriers are doped into the system, its temperature dependence being inherited from anharmonic properties of insulating SrTiO$_3$.

The behavior of center frequency, $\omega_0$, with doping and temperature is summarized in Fig.~\ref{fig:fig4}, where it is displayed against the square-root of the carrier density, $\sqrt{n}$, determined from Hall measurements. If the excitation were a conventional Drude plasmon, for which $\omega_p = \sqrt{n e^2/\epsilon_0 m^*}$, the points would reside on a line that passes through zero, illustrated by the dashed line in Fig.~\ref{fig:fig4}. In fact, the curves tend toward a nonzero intercept at $\sqrt{n}=0$, implying a nonzero excitation frequency even in the absence of carriers. The shift of $\omega_0$ with temperature is present at all dopings, including $x=0$ where no plasmon is present. We conclude that the valence plasmon in SrTi$_{1-x}$Nb$_x$O$_3$ is really a composite excitation with mixed plasmon/phonon character that inherits its unusual temperature dependence from anharmonic properties of the insulator. 

These measurements present a challenge to theory. In the simplest model of a doped semiconductor with a Fr\"{o}hlich electron-phonon coupling, one would expect the dielectric function to exhibit identifiable, coupled phonon-plasmon modes that become overdamped only at momenta higher than the electron-hole continuum \cite{Barker1966}. Such plasmon-phonon interference effects have been observed in CdS and GaN \cite{10.1103/physrevb.3.1295, 10.1088/0953-8984/21/17/174204}. In the case of SrTi$_{1-x}$Nb$_x$O$_3$, the plasmon is never long-lived and never forms a pole that is distinct from the multiphonon continuum present in the insulator. Our measurements therefore reveal an electronic state that is more radically entwined with the lattice. 

The momentum-independent property of the multiphonon-plasmon, observed at all dopings, is a curiosity. A multiphonon continuum is inherently momentum-independent, since it is essentially a density-of-states measurement \cite{Ashcroft76}, so it stands to reason that its lineshape would not depend on $q$ in the insulator. However it is surprising that this property carries over to doped materials in which the excitation has much more plasmon character. 

Additional insight into coupling between plasmon effects and the lattice mix might be obtained by applying two-dimensional optical spectroscopy \cite{Mukamel1995}. Such methods have been successful in disentangling vibrational modes in molecular systems \cite{cundiff2013optical}. Such measurements on SrTi$_{1-x}$Nb$_x$O$_3$ using IR and THz radiation could reveal coupling between different modes, discriminate between homogeneous and inhomogeneous broadening, and reveal the fundamental relaxation and decoherence lifetimes of collective excitations \cite{Wan2019, Choi2020, Li2021, Gerken2022,Lu2016,Lu2017,mahmood2021}.

In summary, we used M-EELS to investigate the IR plasmon in SrTi$_{1-x}$Nb$_x$O$_3$.
We find that the feature previously identified as a free-carrier plasmon \cite{Bi2006, Gervais1993, VanMechelen2008, Collignon2020, Eagles1996, Li2018, Devreese2010, VanDerMarel2011} is also visible in the insulator, though only at $q \neq$ 0, where it may be identified instead as a multiphonon background arising from lattice anharmonicity \cite{Ashcroft76}. This anharmonic background exhibits the same temperature dependence in the insulator as the IR plasmon in doped materials.
Adding carriers by Nb doping increases the spectral weight in this background and draws it to $q=0$ where, at sufficient doping, it becomes visible in IR experiments. We conclude that the IR plasmon in SrTi$_{1-x}$Nb$_x$O$_3$ is not a simple, free-carrier mode, but a composite excitation with mixed plasmon- and multiphonon- character that inherits its anomalous properties from lattice anharmonicity in the insulator ($x=0$).

\nocite{Chen-2019, mcmillan-1976, Nattermann-1997, Landau-QM, Joe-2014, Hoffman-2002, Kivelson-2002, Yu-2019, Kalisky2010,Xiao2012}

We thank Dirk van der Marel and Alexey Kuzmenko for helpful discussions. This work was supported by the Center for Quantum Sensing and Quantum Materials, a DOE Energy Frontier Research Center, under award DE-SC0021238. P.A. acknowledges support from the EPiQS program of the Gordon and Betty Moore Foundation, grant GBMF9452. M.M. acknowledges support from the Alexander von Humboldt foundation.


\bibliography{STObib}

\end{document}